\documentclass[a4paper,10pt,twoside]{cpc-hepnp}

\usepackage{multicol}
\usepackage{graphicx}
\usepackage{booktabs}
\usepackage{amssymb,bm,mathrsfs,bbm,amscd}
\usepackage[tbtags]{amsmath}
\usepackage{lastpage}
\usepackage{multirow}
\usepackage{color}
\usepackage{ulem}

\newcommand{\delete}{\bgroup\markoverwith{\textcolor{red}{\rule[0.5ex]{2pt}{1pt}}}\ULon}

\begin{document}

\fancyhead[c]{\small Chinese Physics C~~~Vol. **, No. * (2017)
******} \fancyfoot[C]{\small ******-\thepage}

\footnotetext[0]{Received ** ***** 2017}

\title{Relativistic interpretation on the nature of nuclear tensor force\thanks{Supported by the National Natural Science Foundation of China under Grant Nos. 11375076 and 11675065, and the Fundamental Research Funds for the Central Universities under Grant No. lzujbky-2016-30}}

\author{%
      Yao Yao Zong$^{1}$
\quad Bao Yuan Sun$^{1;1)}$\email{sunby@lzu.edu.cn}%
}
\maketitle

\address{%
$^1$ School of Nuclear Science and Technology, Lanzhou University, Lanzhou 730000, China\\
}

\begin{abstract}
The spin-dependent nature of the nuclear tensor force is studied in details within the relativistic Hartree-Fock approach. The relativistic formalism for the tensor force is supplemented with an additional Lorentz-invariant tensor formalism in $\sigma$-scalar channel, so as to take into account almost fully the nature of the tensor force brought about by the Fock diagrams in realistic nuclei. Specifically, the tensor sum rules are tested for the spin and pseudo-spin partners with/without nodes, to further understand the tensor force nature within relativistic model. It is shown that the interference between two components of nucleon spinors brings distinct violations on the tensor sum rules in realistic nuclei , which is mainly due to the opposite sign on $\kappa$ quantities of the upper and lower components as well as the nodal difference. Even though, the sum rules can be precisely reproduced if taking the same radial wave functions for the spin/pseudo-spin partners in addition to neglecting the lower/upper components, revealing clearly the nature of tensor force.
\end{abstract}



\footnotetext[0]{\hspace*{-3mm}\raisebox{0.3ex}{$\scriptstyle\copyright$}2017
Chinese Physical Society and the Institute of High Energy Physics
of the Chinese Academy of Sciences and the Institute
of Modern Physics of the Chinese Academy of Sciences and IOP Publishing Ltd}%

\begin{multicols}{2}

\section{Introduction}

In nuclear physics, nuclear force that binds protons and neutrons into atomic nucleus is one of the most important issues, and lots attempts have been devoted to explaining the nature of nuclear force. Nowadays, the meson exchange diagram of nuclear force proposed by Yukawa is still one of the most successful attempts \cite{Yukawa1935}. In past years, the worldwide construction of new generation of the radioactive ion beam facilities has greatly promoted the development of the field and a set of novel phenomena have been discovered in exotic nuclei. It brings a series of challenges and opportunities for nuclear physics, especially in understanding the nature of nuclear force. As one of the typical examples, the non-central tensor force has drawn considerable attentions due to its characteristic spin-dependent nature \cite{Otsuka2005PRL95.232502},  which plays an essential role in determining the nuclear shell evolutions \cite{Otsuka2006PRL97.162501, Brown2006PRC74.061303, Colo2007PLB646.227, Lesinski2007PRC76.014312, Zou2008PRC77.014314, Long2008EPL82.12001,Long2009PLB680.428, Bender2009PRC80.064302, Otsuka2010PRL104.012501, Moreno2010PRC81.064327, Wang2011PRC83.054305, Wang2011PRC84.044333, Dong2011PRC84.014303, Wang2013PRC87.047301, Shen2017arXiv1709.06289}, nuclear isospin excitations and $\beta$-decays \cite{Bai2009PLB675.28, Bai2010PRL105.072501, Cao2009PRC80.064304, Anguiano2011PRC83.064306, Co2012PRC85.034323, Minato2013PRL110.122501}, as well as the nuclear matter properties \cite{Jiang2015PRC91.025802}.

Based on the meson exchange diagram of nuclear force, the relativistic description of nuclear structure properties has achieved great progresses in collaborating with the spirit of the density functional theory, namely the famous covariant density functional (CDF) theory (see Ref. \cite{Meng2016WS} and references therein). During the past decades, the relativistic mean field (RMF) theory, the CDF theory without Fock terms, has been paid more and more attentions due to its successful description of many nuclear phenomena in both stable and exotic nuclei \cite{Reinhard1989RPP52.439, Serot1986ANP16.1, Ring1996PPNP37.193, Lalazissis1997PRC55.540, Serot1997IJPE06.515, Bender2003RMP75.121, Zhou2003PRL91.262501, Long2004PRC69.034319, Meng2006PPNP57.470, Meng2006PRC73.037303, Meng2015JPG42.093101, Zhao2010PRC82.054319}. With the covariant frame, mainly represented as large scalar and vector fields of the order of a few hundred MeV, the RMF theory can provide self-consistent description for the spin-orbit (SO) couplings, an important ingredient of nuclear force. While limited by the RMF approach itself, the important degrees of freedom in the meson exchange diagram, such as the $\pi$- and $\rho$-tensor couplings, are missing, and specifically the important ingredient of nuclear force --- tensor force, arising from the $\pi$ exchange and $\rho$-tensor coupling, cannot be efficiently taken into account. Notice the fact that the $\pi$- and $\rho$-tensor couplings, as well as the tensor force, can be considered only (or mainly) with the presence of Fock terms which are in general ignored in the RMF scheme for simplicity.

Because of the complexity induced by Fock terms and the limitation of computer power, it remains as a long standing problem to provide an appropriate quantitative description of nuclear structure properties under the relativistic Hartree-Fock (RHF) approach \cite{Bouyssy1987PRC36.380, Bernardos1993PRC48.2665, Shi1995PRC52.144, Marcos2004JPG30.703, Shen2016CPL33.102103}. Until ten years ago, a new RHF approach, namely the density-dependent relativistic Hartree-Fock (DDRHF) theory \cite{Long2006PLB640.150, Long2007PRC76.034314, Long2010PRC81.024308}, also referred as the CDF theory with Fock terms, was developed in collaboration with the density-dependent meson-nucleon coupling, and a quantitative description of nuclear structure properties was achieved with comparable accuracy to the standard CDF models. Due to the Lorentz covariance, the RHF approach maintains the advantages as well as the RMF one, i.e., the self-consistent treatment of the spin-orbit coupling. Moreover, the presence of Fock terms has brought significant improvements in describing nuclear properties, such as self-consistent description of shell evolution \cite{Long2008EPL82.12001, Long2009PLB680.428, Wang2013PRC87.047301, Li2014PLB732.169, Li2016PLB753.97}, better preserved pseudo-spin and spin symmetries \cite{Long2007PRC76.034314, Long2006PLB639.242, Long2010PRC81.031302, Liang2010EPJA44.119, Li2016PRC93.054312}, and fully self-consistent treatment of nuclear isospin excitation \cite{Liang2008PRL101.122502, Liang2009PRC79.064316, Liang2012PRC85.064302} and decay modes \cite{Niu2013PLB723.172, Niu2017PRC95.044301}. Besides, the Fock terms also present distinct contributions to nuclear symmetry energy \cite{Sun2008PRC78.065805, Long2012PRC85.025806, Zhao2015JPG42.095101}.

Recently, the analysis within the DDRHF theory show that the Fock terms of the meson-nucleon couplings represent distinct spin dependence \cite{Jiang2015PRC91.034326}, a characteristic nature of tensor force \cite{Otsuka2005PRL95.232502}. It was then recognized that the Fock diagrams of the meson-nucleon couplings can take the important ingredient of nuclear force --- the tensor force into account naturally \cite{Jiang2015PRC91.034326}. Particularly, more remarkable tensor effects are found in the Fock terms of the isoscalar $\sigma$-S and $\omega$-V couplings, rather than the isovector $\rho$-V, $\rho$-T and $\pi$-PV couplings. In Ref. \cite{Jiang2015PRC91.034326}, a series of relativistic formalism have been proposed for the tensor force components in the Fock diagrams of various meson-nucleons couplings, and the self-consistent tensor effects were analysed for nuclear ground states and nuclear matter with the proposed relativistic formalism \cite{Jiang2015PRC91.025802, Jiang2015PRC91.034326}. Without introducing any additional free parameter, the spin-dependent feature brought about by the Fock terms can be interpreted almost completely by the proposed relativistic formalism. In addition, the reduction of the kinetic part of symmetry energy at the supranuclear density region in the DDRHF theory can be regarded partly as the effect of the nuclear tensor force \cite{Zhao2015JPG42.095101}.

Notice that the tensor force, as derived from shell model calculation \cite{Otsuka2005PRL95.232502}, shall fulfill some specific sum rules [see Eqs. (\ref{equ:sumrule}) and (\ref{equ:sumrule2})] quantitatively owing to its spin-dependent feature. Conceptually, the sum rules were verified in Ref. \cite{Jiang2015PRC91.034326} under the assumption of neglecting the lower components of Dirac spinors and taking same radial wave functions for the spin partner $j_\pm = l\pm 1/2$ states. Since the lower and upper components of Dirac spinors are of different angular momenta, leading to opposite parity, the sum rules could be violated distinctly with complete form of the Dirac spinor in realistic nuclei. To better understand the nature of the tensor force, it is worthwhile to test the sum rule in the realistic nuclei with the relativistic formalism of tensor forces, and to reveal the relativistic effect brought about by the lower component of Dirac spinor. The contents are organized as follows. In Sec. 2, the relativistic formalism for the tensor force components in the Fock diagrams of meson-nucleon couplings are recalled and the supplementation to the one in $\sigma$-S coupling channel is presented. In Sec. 3 the verification on the sum rules are performed by taking the spin/pseudo-spin parters in $^{48}$Ca, $^{90}$Zr and $^{208}$Pb as the examples to understand the nature of the tensor force, and the contributions from the lower/upper components are discussed in details. Finally, a brief summary is given in Sec. 4.

\section{Supplementation of the relativistic formalism for tensor force components in $\sigma$-S channel}

For completeness, the relativistic formalism for the tensor force components in the Fock terms of $\sigma$-S, $\omega$-V, $\rho$-V, $\rho$-T and $\pi$-PV couplings are recalled as follows,
\begin{align}
  \mathscr H_{\sigma\text{-S}}^{T_1} = & -\frac{1}{2}\cdot \frac{1}{2}\Big[\frac{g_\sigma}{m_\sigma} \bar\psi\gamma_0\Sigma_\mu\psi\Big]_1 \Big[\frac{g_\sigma}{m_\sigma} \bar\psi\gamma_0\Sigma_\nu\psi\Big]_2 D_{\sigma\text{-S}}^{T,\ \mu\nu}(1,2),\label{equ:sigma}\\
  \mathscr H_{\omega\text{-V}}^T = & +\frac{1}{2}\cdot \frac{1}{2}\Big[\frac{g_\omega}{m_\omega} \bar\psi\gamma_\lambda\gamma_0\Sigma_\mu\psi\Big]_1 \Big[\frac{g_\omega}{m_\omega} \bar\psi\gamma_\delta\gamma_0\Sigma_\nu\psi\Big]_2 D_{\omega\text{-V}}^{T,\ \mu\nu\lambda\delta}(1,2),\label{equ:omega}\\
  \mathscr H_{\rho\text{-T}}^T = & + \frac{1}{2}\Big[\frac{f_\rho}{2M}\bar\psi\sigma_{\lambda\mu}\vec{\tau}\psi\Big]_1\cdot \Big[\frac{f_\rho}{2M}\bar\psi\sigma_{\delta\nu}\vec{\tau}\psi\Big]_2 D_{\rho\text{-T}}^{T,\ \mu\nu\lambda\delta}(1,2),\label{equ:rho-T}\\
  \mathscr H_{\pi\text{-PV}}^T = & - \frac{1}{2}\Big[\frac{f_\pi}{m_\pi}\bar\psi\gamma_0\Sigma_\mu\vec{\tau}\psi\Big]_1\cdot \Big[\frac{f_\pi}{m_\pi}\bar\psi\gamma_0\Sigma_{\nu}\vec{\tau}\psi\Big]_2 D_{\pi\text{-PV}}^{T,\ \mu\nu}(1,2),\label{equ:pio-PV}
\end{align}
where the relativistic spin operator $\Sigma^\mu = \big(\gamma^5,\bm\Sigma)$, $M$ is the nucleon mass, and $\vec{\tau}$ denotes the isospin operator of nucleon ($\psi$). The propagator terms $D^T$ read as,
\begin{align}
  D_\phi^{T,\ \mu\nu}(1,2) = & \Big[\partial^\mu(1)\partial^\nu(2) - \frac{1}{3}g^{\mu\nu} m_\phi^2\Big] D_\phi(1,2)\nonumber\\ &\hspace{2em} + \frac{1}{3}g^{\mu\nu} \delta(x_1-x_2),\\
  D_{\phi'}^{T,\ \mu\nu\lambda\delta}(1,2) = & \partial^\mu(1)\partial^\nu(2)g^{\lambda\delta} D_{\phi'}(1,2) \nonumber\\ &\hspace{2em} - \frac{1}{3}\Big(g^{\mu\nu}g^{\lambda\delta} - \frac{1}{3} g^{\mu\lambda}g^{\nu\delta}\Big) m_{\phi'}^2 D_{\phi'}(1,2)\nonumber\\ &\hspace{2em} + \frac{1}{3}\Big(g^{\mu\nu}g^{\lambda\delta} - \frac{1}{3} g^{\mu\lambda}g^{\nu\delta}\Big)\delta(x_1-x_2),\label{equ:Dphip}
\end{align}
where $\phi$ stands for $\sigma$-S and $\pi$-PV channels, and $\phi'$ represents $\omega$-V and $\rho$-T channels. Here to distinguish with the following one, we use $\mathscr H_{\sigma\text{-S}}^{T_1}$ to denote the relativistic formalism for tensor force components in $\sigma$-S channel proposed by Ref. \cite{Jiang2015PRC91.034326}. For the $\rho$-V channel, corresponding formalism $\mathscr H_{\rho\text{-V}}^T$ can be obtained simply by replacing $m_\omega (g_\omega)$ in Eqs. (\ref{equ:omega}) and (\ref{equ:Dphip}) by $m_\rho (g_\rho)$ and inserting the isospin operator $\vec{\tau}$ in the interacting index.

As the prior study on the tensor effects brought about by the Fock terms, the lower components of Dirac spinors were dropped and the spin partners $j_\pm = l\pm 1/2$ were assumed to share the same radial wave functions \cite{Jiang2015PRC91.034326}. Naturally, it would be interesting to study the tensor effects with complete form of the relativistic wave functions, namely for realistic nuclei. Utilizing the full Dirac spinors determined by the self-consistent calculations with RHF-PKA1 model, we calculate the contributions from various single-particle orbits $j'$ to the SO splitting of the spin-partner states $j_\pm$, defined as
\begin{align}
\Delta E_{\text{SO}} \equiv V_{j_-j'}-V_{j_+j'},
\end{align}
where $V_{jj'}$ denotes the interaction matrix element in the single-particle states $j$ and $j'$. In fact, not only the contributions from the direct terms but also those from the exchange parts of single-particle potentials give rise to the SO splitting in nuclear single-particle spectra. However, it is unveiled that the spin-dependent feature of the contributions $\Delta E_{\text{SO}}$ from the nucleon-nucleon interactions, namely the difference in its values between the spin-partner states $j'_\pm=l\pm 1/2$, is dominated by the Fock diagrams, particularly via the isoscalar meson-nucleon coupling channels \cite{Jiang2015PRC91.034326}. Such a spin dependence of $\Delta E_{\text{SO}}$ is then well explained by introducing a series of Lorentz-invariant tensor formalism (Eqs. (\ref{equ:sigma}-\ref{equ:pio-PV})), and the corresponding tensor interaction matrix element reads as
\begin{align}
V^T_{jj'}=\bar f_{j}(\bm r_1) \bar f_{j'}(\bm r_2)\Gamma^T_{1,2}f_{j}(\bm r_2)f_{j'}(\bm r_1),
\end{align}
with $\Gamma^T_{1,2}$ the interaction vertices of tensor force and the nucleon Dirac spinor
\begin{align}
f_\alpha(\bm r)=\frac{1}{r}\left(
\begin{array}{c}
iG_a(r)\\
F_a(r)\hat{\bm\sigma}\cdot\hat{\bm r}
\end{array}
\right)
\mathscr{Y}_a(\hat{\bm r})\chi_{\frac{1}{2}}(q_a),
\end{align}
where $\mathscr{Y}_a(\hat{\bm r})$ is the spinor spherical harmonic, and $\chi_{\frac{1}{2}}(q_a)$ the isospinor. $G_a$ and $F_a$ correspond to the upper and lower components of the radial wave function, respectively. Thus, $V^T_{jj'}$ can be divided further with respect to the $G_a$ and $F_a$ components of the states $j$ and $j'$, namely,
\begin{align}\label{equ:VT4}
    V^T_{jj'} = V^{T}_{G_jG_{j'}} + V^{T}_{G_jF_{j'}} + V^{T}_{F_jG_{j'}} + V^{T}_{F_jF_{j'}}.
\end{align}

Figure \ref{fig:DSO-so} shows the contributions to $\Delta E_{\text{SO}}$ only from the Fock diagrams, giving the total, the tensor parts and the central ones, respectively, taking the nodeless neutron ($\nu$) orbits in $^{48}$Ca as examples. From Figs. \ref{fig:DSO-so}(b) and \ref{fig:DSO-so}(c), it is found that the spin-dependent feature of $\Delta E_{\text{SO}}$ in Fock terms can be almost fully interpreted by the relativistic formalism for the tensor force components in the $\omega$-V channel [i.e., Eq. (\ref{equ:omega})], while it seems to be overestimated in $\sigma$-S coupling channel, implying that the spin-dependent feature introduced by the $\sigma$-S coupling cannot be exactly explained by Eq. (\ref{equ:sigma}).

\begin{center}
\includegraphics[width=8cm]{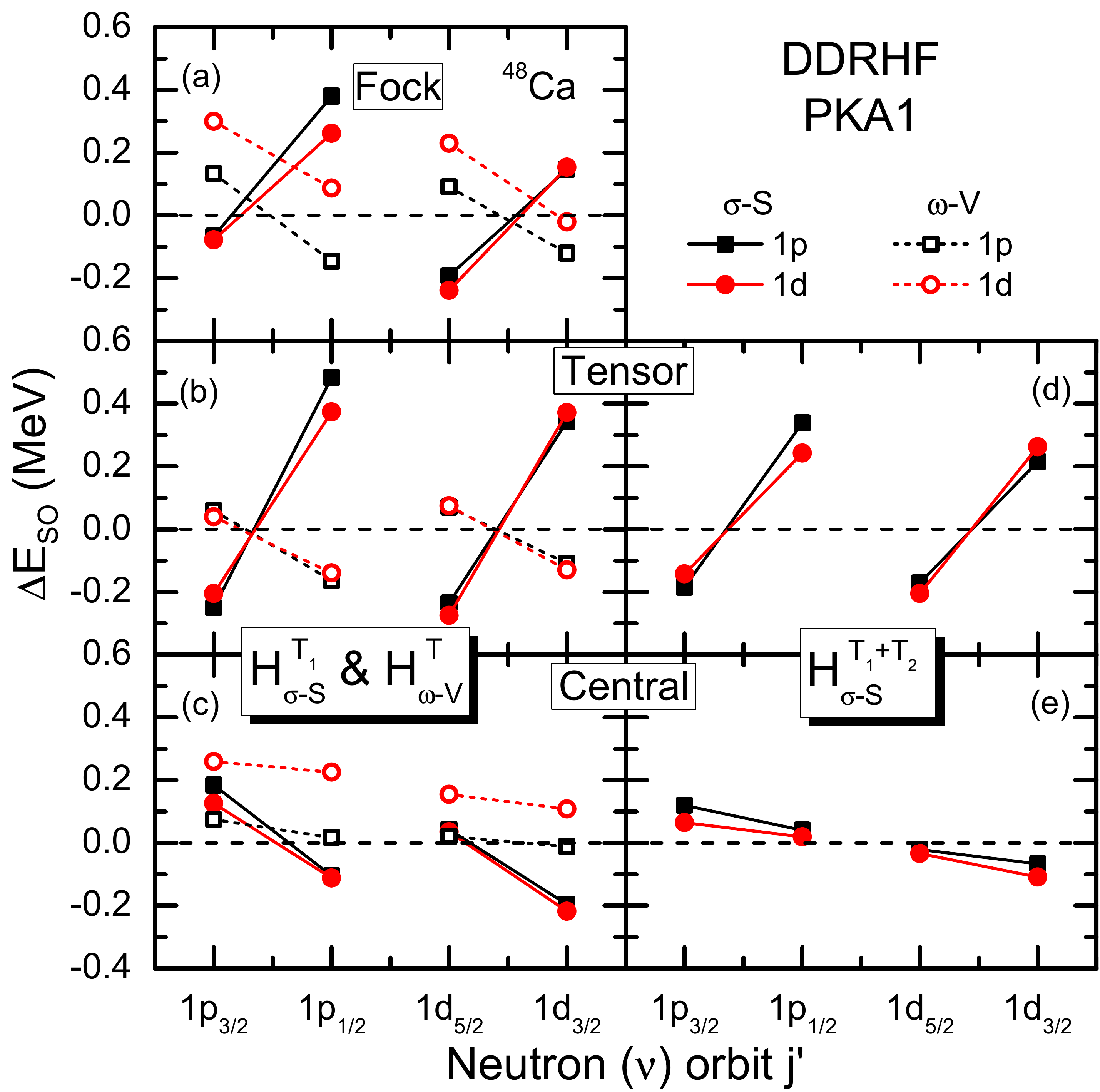}
\figcaption{\label{fig:DSO-so}   (Color Online) Contributions to the spin-orbit (SO) splittings $\Delta E_{\text{SO}}=V_{j_-j'}-V_{j_+j'}$ (MeV): the Fock terms [plot (a)], the tensor parts [plots (b) and (d)] and the central ones [plot (c) and (e)]. In plots (d-e), the supplemented tensor terms $\mathscr H_{\sigma\text{-S}}^{T_2}$ are taken into account. The result are extracted from the calculations of DDRHF functional PKA1 for nodeless neutron ($\nu$) orbits in $^{48}$Ca, and full Dirac spinors are utilized in calculating the interaction matrix element $V_{jj'}$. }
\end{center}

It should be mentioned that the $\pi$-PV coupling is originally in form of Lorentz tensor couplings of rank 2 \cite{Bouyssy1987PRC36.380, Long2006PLB640.150}, and it is proved that the spin-dependent feature introduced by the $\pi$-PV coupling can be fully explained indeed by $\mathscr H_{\pi\text{-PV}}^T$, even without introducing the approximation on the radial wave functions. In fact, from the relativistic formalism of the tensor forces in Eqs. (\ref{equ:sigma}-\ref{equ:pio-PV}), one can find that all the formalism certainly represent as the relativistic-type (Lorentz) tensor couplings but taking various ranks, formally at rank 4 for the $\omega$-V and $\rho$-T channels, and for the later it is actually at rank 3, and the ones for the $\sigma$-S and $\pi$-PV channels are in form of rank 2. In order to avoid overestimation on the tensor effects involved by the Fock terms of $\sigma$-S coupling [see Figs. \ref{fig:DSO-so}(b) and \ref{fig:DSO-so}(c)], higher rank of Lorentz tensor coupling than Eq. (\ref{equ:sigma}) could be considered as an supplementation. The relativistic formalism of tensor force at rank 3, similar to $\mathscr H_{\rho\text{-T}}^T$, is the first choice naturally, i.e.,
\begin{align}
\mathscr H^{T_2}_{\sigma\text{-S}}=+\frac{1}{4}\cdot\frac{1}{9} \Big[\frac{g_\sigma}{m_\sigma}\bar\psi\sigma_{\lambda\mu}\psi\Big]_1 \cdot \Big[\frac{g_\sigma}{m_\sigma}\bar\psi\sigma_{\delta\nu}\psi\Big]_2 D^{T,\mu\nu\lambda\delta}_{\sigma\text{-S}}(1,2), \label{equ:sigma2}
\end{align}
where the propagator term $D^{T,\mu\nu\lambda\delta}_{\sigma\text{-S}}$ corresponds to Eq. (\ref{equ:Dphip}) with $\phi'=\sigma$-S. As shown in Figs. \ref{fig:DSO-so}(d) and \ref{fig:DSO-so}(e), the supplemented relativistic formalism for the tensor force components in $\sigma$-S channel, i.e., $\mathscr H_{\sigma\text{-S}}^{T_1+T_2}$, improves remarkably the description of the tensor nature involved by the Fock terms of $\sigma$-S coupling, leading to a clear damping on the spin dependence of the central part (Figs. \ref{fig:DSO-so}(e)) in comparison with that in the case of $\mathscr H_{\sigma\text{-S}}^{T_1}$ (Figs. \ref{fig:DSO-so}(c)). It is worthwhile to mention again that the full Dirac spinors determined self-consistently with RHF-PKA1 model are utilized in calculating $\Delta E_{\rm SO}$, different from Ref. \cite{Jiang2015PRC91.034326}. For the other coupling channels, namely $\omega$-V, $\rho$-V, $\rho$-T and $\pi$-PV, it has been checked that the existing relativistic formalism of the tensor forces [i.e., Eqs. (\ref{equ:omega}-\ref{equ:pio-PV})] can describe fully the tensor effects involved by the relevant Fock diagrams even with the complete Dirac spinors.

As derived from the shell model calculations \cite{Otsuka2005PRL95.232502}, the tensor force shall fulfill the following sum rule,
\begin{align}
\label{equ:sumrule}
    V_{\text{sum}} \equiv
    \hat j_+^2 V^T_{j_+j'}+\hat j_-^2 V^T_{j_-j'} = 0
\end{align}
where $\hat j^2 = 2j+1$. As mentioned in Refs. \cite{Otsuka2005PRL95.232502, Jiang2015PRC91.034326}, one has to choose the same radial wave functions for the spin partner states $j_\pm$ to reproduce the sum rule exactly. In fact, the sum rule is often taken as the identity of the tensor force, which reveals clearly its spin-dependent nature. It is then worthwhile to see how precisely the sum rule is fulfilled by the relativistic formalism of tensor forces, even without introducing the approximation on the wave functions, for instance in realistic nucleus. Particularly, as indicated by the supplemented relativistic formalism $\mathscr H_{\sigma\text{-S}}^{T_1+T_2}$ for the $\sigma$-S channels, it is also valuable to verify the role of the lower component of nucleon spinor, from which the relativistic effects are expected to be revealed.

\begin{center}
\tabcaption{ \label{tab:sumrule-sig}  Interaction matrix elements $V_{j_\pm j'}^T$ ($10^{-1}$ MeV) between the spin partner states, namely the nodeless neutron orbits $p$, $d$ and $f$ of $^{48}$Ca, for the tensor force components in the Fock diagram of the $\sigma$-S couplings. The 2$^{\text{nd}}$-7$^{\text{th}}$ rows shows the results calculated with the radial wave functions determined by the self-consistent calculations of DDRHF with PKA1. For the results in the 8$^{\text{th}}$-13$^{\text{th}}$ rows the lower components in both the $j_\pm$ and the $j'_\pm$ orbits are omitted, and for ones in the 14$^{\text{th}}$-19$^{\text{th}}$ rows the $j_\pm$ orbits, as well as $j'_\pm$ orbits, share the same radial wave functions in addition to neglecting the lower components. }
\footnotesize
 \begin{tabular}{c|rrrrr}\hline\hline
\multirow{2}{*}{$j_\pm$} & \multicolumn{5}{c}{$j'$} \\ \cline{2-6}
              &$\nu1p_{3/2}$&$\nu 1p_{1/2}$&$\nu1d_{5/2}$&$\nu1d_{3/2}$&$\nu1f_{7/2}$\\\hline
$\nu1p_{3/2}$ &       0.463 & $-$1.391     &      0.401  &   $-$1.031  &      0.259  \\
$\nu1p_{1/2}$ &    $-$1.391 &    2.000     &   $-$1.310  &      1.113  &   $-$0.884  \\
$V_{\rm{sum}}$&    $-$0.929 & $-$1.563     &   $-$1.015  &   $-$1.900  &   $-$0.731  \\\hline
$\nu1d_{5/2}$ &       0.401 & $-$1.310     &      0.691  &   $-$1.365  &      0.597  \\
$\nu1d_{3/2}$ &    $-$1.031 &    1.113     &   $-$1.365  &      1.261  &   $-$1.189  \\
$V_{\rm{sum}}$&    $-$1.719 & $-$3.409     &   $-$1.312  &   $-$3.146  &   $-$1.175  \\\hline\hline
$\nu1p_{3/2}$ &       0.635 & $-$1.335     &      0.622  &   $-$0.963  &      0.433  \\
$\nu1p_{1/2}$ &    $-$1.335 &    2.818     &   $-$1.263  &      1.968  &   $-$0.848  \\
$V_{\rm{sum}}$&    $-$0.132 &    0.298     &   $-$0.038  &      0.087  &      0.036  \\\hline
$\nu1d_{5/2}$ &       0.622 & $-$1.263     &      0.875  &   $-$1.295  &      0.795  \\
$\nu1d_{3/2}$ &    $-$0.963 &    1.968     &   $-$1.295  &      1.932  &   $-$1.142  \\
$V_{\rm{sum}}$&    $-$0.118 &    0.295     &      0.072  &   $-$0.041  &      0.204  \\\hline\hline
$\nu1p_{3/2}$ &       0.635 & $-$1.269     &      0.622  &   $-$0.933  &      0.433  \\
$\nu1p_{1/2}$ &    $-$1.269 &    2.538     &   $-$1.244  &      1.866  &   $-$0.866  \\
$V_{\rm{sum}}$&    $-$0.000 &    0.000     &      0.000  &      0.000  &      0.000  \\\hline
$\nu1d_{5/2}$ &       0.622 & $-$1.244     &      0.875  &   $-$1.313  &      0.795  \\
$\nu1d_{3/2}$ &    $-$0.933 &    1.866     &   $-$1.313  &      1.969  &   $-$1.193  \\
$V_{\rm{sum}}$&       0.000 &    0.000     &   $-$0.000  &   $-$0.000  &      0.000  \\\hline\hline
 \end{tabular}
\end{center}

\section{Verification for Tensor Sum Rule }\label{sec:sumrule}

In this section, the tensor sum rule  is verified under the relativistic Hartree-Fock (RHF) approach with the functional PKA1. In order to provide a detailed understanding on the sum rule, the spin-unsaturated magic system, namely $^{48}$Ca, $^{90}$Zr and $^{208}$Pb, are taken as the examples in the following and the discussion focuses on the neutron spin partner states without/with nodes, and the pseudo-spin partners, respectively.

\subsection{Sum Rule for Spin Partner States}

Taking the nodeless neutron orbits $p$, $d$ and $f$ of $^{48}$Ca as the examples, Table \ref{tab:sumrule-sig} shows the tensor interaction matrix elements $V_{j_\pm j'}^T$ (in units of $10^{-1}$ MeV), described by the supplemented relativistic formalism for the nuclear tensor force components in the Fock diagram of $\sigma$-S coupling. The 2$^{\text{nd}}$ to the 7$^{\text{th}}$ rows shows the results calculated with the full nucleon spinors. It is found that there exist rather distinct deviations from the sum rule, with the $V_{\rm sum}$ values comparable to the interaction matrix elements themselves. If one neglects the lower components in nucleon spinors, i.e., the results in the 8$^{\text{th}}$ to the 13$^{\text{th}}$ rows, it can be seen that the sum rule is properly fulfilled with the relative deviations of few percents. If taking the same assumption as in Ref. \cite{Jiang2015PRC91.034326}, i.e., the $j_+$ and $j_-$ orbits, as well as the $j'_+$ and $j'_-$ orbits, share the same radial wave functions in addition to neglecting the lower components, the sum rule is reproduced precisely with negligible errors ($V_{\rm sum} \lesssim 10^{-6}$ MeV).

Similar to Table \ref{tab:sumrule-sig}, Table \ref{tab:sumrule-ome} shows the tensor interaction matrix elements described by the relativistic formalism for the nuclear tensor force components in the Fock diagram of $\omega$-V coupling, and similar systematics in reproducing the sum rule are found. In fact, besides the sum rule in Eq. (\ref{equ:sumrule}), the interaction matrix elements of the tensor force shall fulfill the following relations as well,
\begin{align}\label{equ:sumrule2}
  \hat j_+^2\hat j_+'^2V_{j_+j'_+}^T - \hat j_-^2\hat j_-'^2 V_{j_-j'_-}^T = & 0, &
  \hat j_+^2\hat j_-'^2V_{j_+j'_-}^T - \hat j_-^2\hat j_+'^2 V_{j_-j'_+}^T = & 0,
\end{align}
which are precisely reproduced by the proposed relativistic formalism, if both $j_\pm$ and $j'_\pm$ states share the same radial wave functions in addition to neglecting the lower components (see the last six rows in Tables \ref{tab:sumrule-sig} and \ref{tab:sumrule-ome}). Moreover, comparing the results of $V_{jj'}^T$ with full nucleon spinors and the ones neglecting the lower components in Tables \ref{tab:sumrule-sig} and \ref{tab:sumrule-ome}, it is seen that the influence induced by the inclusion of the lower components is more remarkable for the repulsive-type tensor interaction matrix elements, i.e., $V_{j_+j'_+}^T$ and $V_{j_- j_-'}^T$ in Table \ref{tab:sumrule-sig}, and $V_{j_-j'_+}^T$ and $V_{j_+ j_-'}^T$ in Table \ref{tab:sumrule-ome}, rather than for the attractive-type ones.

\begin{center}
\tabcaption{ \label{tab:sumrule-ome}  Similar to Table \ref{tab:sumrule-sig}, but for the tensor force components in the Fock diagram of the $\omega$-V couplings.}
\footnotesize
\begin{tabular}{c|rrrrr}\hline\hline
\multirow{2}{*}{$j_\pm$} & \multicolumn{5}{c}{$j'$} \\ \cline{2-6}
              &$\nu1p_{3/2}$&$\nu1p_{1/2}$&$\nu1d_{5/2}$&$\nu1d_{3/2}$& $\nu1f_{7/2}$\\\hline
$\nu1p_{3/2}$ &    $-$0.247 &       0.353 &   $-$0.250  &      0.150  &    $-$0.181  \\
$\nu1p_{1/2}$ &       0.353 &    $-$1.278 &      0.451  &   $-$0.935  &       0.341  \\
$V_{\rm{sum}}$&    $-$0.282 &    $-$1.144 &   $-$0.101  &   $-$1.272  &    $-$0.042  \\\hline
$\nu1d_{5/2}$ &    $-$0.250 &       0.451 &   $-$0.361  &      0.389  &    $-$0.339  \\
$\nu1d_{3/2}$ &       0.150 &    $-$0.935 &      0.389  &   $-$0.904  &       0.410  \\
$V_{\rm{sum}}$&    $-$0.904 &    $-$1.036 &   $-$0.613  &   $-$1.285  &    $-$0.392  \\\hline\hline
$\nu1p_{3/2}$ &    $-$0.256 &       0.540 &   $-$0.265  &      0.412  &    $-$0.194  \\
$\nu1p_{1/2}$ &       0.540 &    $-$1.145 &      0.540  &   $-$0.845  &       0.381  \\
$V_{\rm{sum}}$&       0.057 &    $-$0.128 &      0.020  &   $-$0.044  &    $-$0.014  \\\hline
$\nu1d_{5/2}$ &    $-$0.265 &       0.540 &   $-$0.380  &      0.564  &    $-$0.357  \\
$\nu1d_{3/2}$ &       0.412 &    $-$0.845 &      0.564  &   $-$0.844  &       0.513  \\
$V_{\rm{sum}}$&       0.058 &    $-$0.144 &   $-$0.026  &      0.008  &    $-$0.092  \\\hline\hline
$\nu1p_{3/2}$ &    $-$0.256 &       0.512 &   $-$0.265  &      0.397  &    $-$0.194  \\
$\nu1p_{1/2}$ &       0.512 &    $-$1.024 &      0.530  &   $-$0.794  &       0.388  \\
$V_{\rm{sum}}$&    $-$0.000 &       0.000 &   $-$0.000  &      0.000  &       0.000  \\\hline
$\nu1d_{5/2}$ &    $-$0.265 &       0.530 &   $-$0.380  &      0.570  &    $-$0.357  \\
$\nu1d_{3/2}$ &       0.397 &    $-$0.794 &      0.570  &   $-$0.855  &       0.536  \\
$V_{\rm{sum}}$&    $-$0.000 &    $-$0.000 &      0.000  &      0.000  &       0.000  \\
\hline\hline
 \end{tabular}
\end{center}

In order to understand the distinct difference on the sum rule (Eq. (\ref{equ:sumrule})) with/without the lower components of nucleon spinors in Tables \ref{tab:sumrule-sig} and \ref{tab:sumrule-ome}, Fig. \ref{fig:Ca48wave} shows the radial wave functions for the upper and lower components. It can be seen that the upper radial wave functions $G(r)$ of the spin partner states are nearly identical with each other, which well explains the proper reproduction of the sum rule shown in the mid six rows of Tables \ref{tab:sumrule-sig} and \ref{tab:sumrule-ome}. However, the radial wave functions $F(r)$ of the lower components, which emerge naturally with the relativistic treatment of nucleon field, are quite different for the spin partners. Such difference in $F(r)$ is ascribed mainly to the so-called nodal effect, that is the node numbers of $G(r)$ and $F(r)$ are the same for $j_+$ states but different by one for $j_-$ states. Therefore, although the component $V^{T}_{G_jG_{j'}}$ in Eq. \eqref{equ:VT4} fulfill appropriately the tensor sum rule in Eq. \eqref{equ:sumrule}, those $V_{jj'}^T$ components involving $F(r)$, namely $V^{T}_{G_jF_{j'}}$, $V^{T}_{F_jG_{j'}}$ and $V^{T}_{F_jF_{j'}}$, could break the sum rule impressively.

Despite the remarkable violation of the sum rule with full nucleon spinors, it is still found that the sum rule is in general better fulfilled for the couplings $V_{j_\pm j'_+}^T$ than the ones $V_{j_\pm j'_-}^T$. It can be understood directly by looking through the role of $V^{T}_{G_jF_{j'}}$, since the contributions from the other components in Eq. \eqref{equ:VT4} are similar in both cases. As has been claimed by the nodal effect, the structure of the lower radial wave functions $F_{j'_-}(r)$ in $j'_-$ states is actually more complicated than $F_{j'_+}(r)$ in $j'_+$ states, consequently leading to the sum rule violated further for $V^{T}_{G_{j_\pm}F_{j'_-}}$ than $V^{T}_{G_{j_\pm}F_{j'_+}}$.

\begin{center}
\includegraphics[width=8cm]{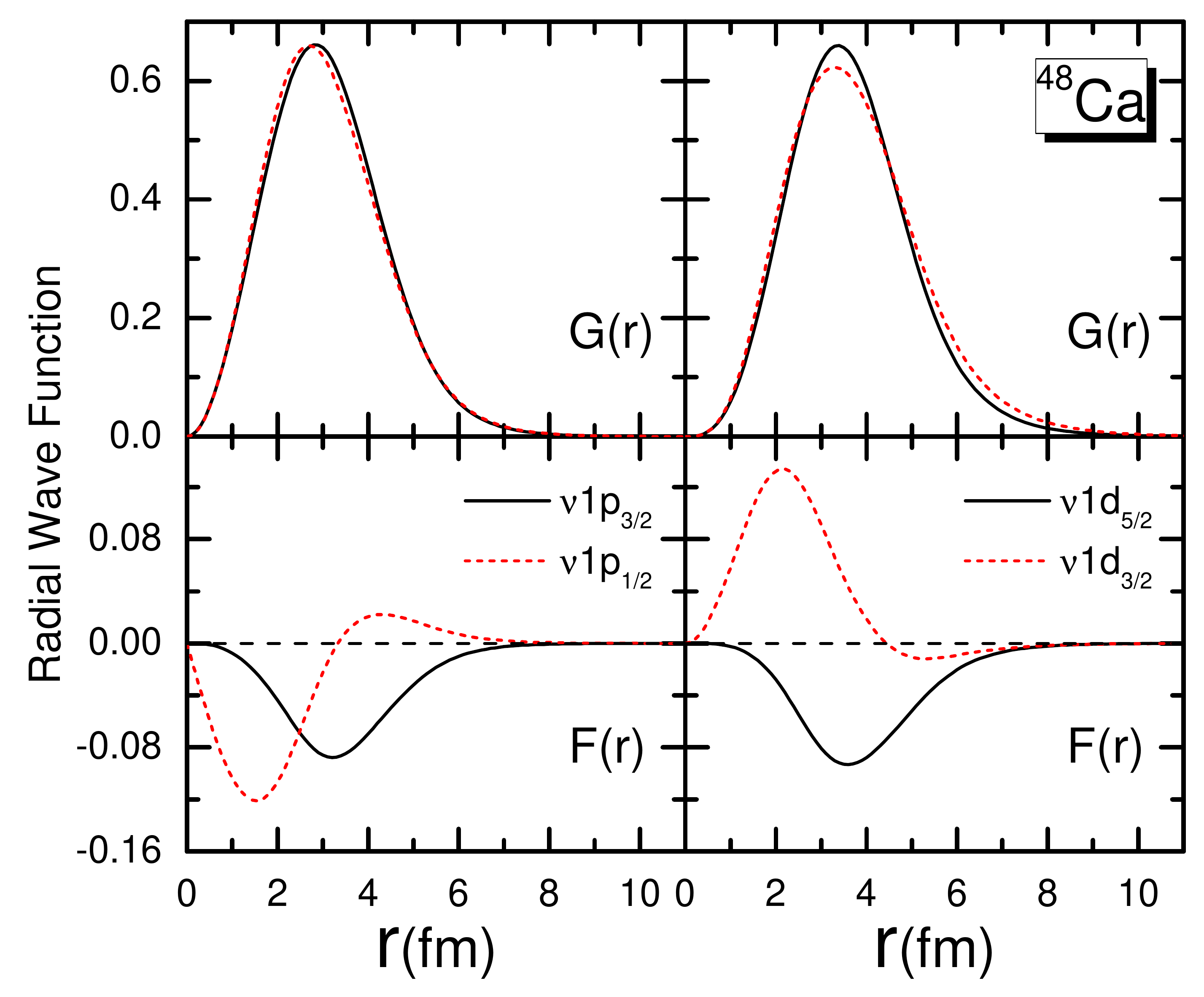}
\figcaption{\label{fig:Ca48wave}   (Color Online) Radial wave functions of the spin partner states $\nu1p$ and $\nu1d$ in $^{48}$Ca, where $G(r)$ denotes the upper components and $F(r)$ corresponds to the lower components. The results are extracted from the calculations of DDRHF with PKA1.}
\end{center}

Besides, the sum rule is checked also for the case of utilizing the full Dirac spinors but assuming that the $j_\pm$ orbits share the same radial wave functions, i.e., the lower components of the spin partner states are set to be the same rather than omitted, which actually can be regarded as eliminating the nodal effect of Dirac spinors. It is found that there still exists distinct violation of the sum rule shown in Eq. \eqref{equ:sumrule}. Hence, the result implies the tensor interaction matrix elements from the lower components of Dirac spinors might correlate with a new tensor sum rule, which could be different from the case of the upper components of Dirac spinors and is deserved to investigate further.

\subsection{Sum Rule for Pseudo-spin Partner States}

In nuclear single-particle spectra, the near degeneration between the single-particle states with quantum numbers ($n$, $l$, $j=l+1/2$) and ($n-1$, $l+2$, $j=l+3/2$) was recognized as the pseudo-spin symmetry (PSS), and the doublet states are referred as the pseudo-spin (PS) partners with newly defined quantum numbers $\tilde n=n-1$, $\tilde l=l+1$, $\tilde j_\pm=\tilde l\pm1/2$ \cite{Arima1969PLB30.517, Hecht1969NPA137.129}. Nowadays the PSS is widely accepted as the relativistic symmetry \cite{Ginocchio1997PRL78.436}, and the quantum number $\tilde{l}$ is nothing but the orbit angular momentum of the lower components of the Dirac spinor. In recent decades, lots of efforts have been devoted to understanding the nature of PSS and the conservation conditions within the relativistic mean field models with/without the Fock terms\cite{Long2009PLB680.428, Long2006PLB640.150, Long2007PRC76.034314, Long2010PRC81.031302, Meng1998PRC58.R628, Meng1999PRC59.154, Meng2006PPNP57.470, Ginocchio2004PRC69.034303, Liang2015PR570.1, Liang2013PRC87.014334, Shen2013PRC88.024311, Chen2003CPL20.358, Lu2012PRL109.072501, Chen2012PRC85.054312, Guo2014PRL112.062502, Guo2005PRC72.054319, Guo2006PRC74.024320}. Similar to the spin partners, it is also worthwhile to check the sum rule for the pseudo-spin partner states with the proposed relativistic formalism of the tensor forces, which could be helpful to understand the role of tensor force in determining the PSS. Correspondingly, the sum rule for pseudo-spin partners can be expressed as,
\begin{align}\label{equ:sumrule3}
\tilde V_{\text{sum}} \equiv &\hat{\tilde j}_+^2 V^T_{\tilde j_+\tilde j'}+\hat{\tilde{j_-^2}} V^T_{\tilde j_-\tilde j'}=0
\end{align}
where $\hat{\tilde j}^2 = 2\tilde j+1$, and $\tilde j_\pm$ denotes the pseudo-spin partners.

Taking the neutron pseudo-spin doublets 1$\tilde p$ and 1$\tilde d$ in $^{90}$Zr as the examples, Table \ref{tab:sumrule-sig-ps} shows the interaction matrix elements $V^T_{\tilde j_\pm \tilde j'}$ between the pseudo-spin partners, calculated with the supplemented relativistic formalism for the tensor force components in the Fock diagram of $\sigma$-S coupling. In Table \ref{tab:sumrule-sig-ps}, the 2$^{\text{nd}}$ to the 7$^{\text{th}}$ rows shows the results (in units of $10^{-1}$ MeV) calculated with the full nucleon spinors. It is found that the sum rule is violated completely, with the $\tilde V_{\rm sum}$ values one order of magnitude larger than the $V^T_{\tilde j_\pm \tilde j'}$ values, and even the spin-dependent characteristic of the tensor forces is not regular as the cases of spin partners anymore. However, if one neglects the upper components of nucleon spinors, which correspond to the results (in units of $10^{-4}$ MeV) in the 8$^{\text{th}}$ to the 13$^{\text{th}}$ rows, it can be seen that the accordance with the sum rule is improved remarkably, while relatively less good than the cases neglecting the lower components of nucleon spinors for spin partners (see the 8-13$^{\text{th}}$ in Tables \ref{tab:sumrule-sig} and \ref{tab:sumrule-ome}). Furthermore, if the $\tilde j_+$ and $\tilde j_-$ orbits, as well as for $\tilde j'_+$ and $\tilde j'_-$ orbits, share the same radial wave functions, i.e., the results (in units of $10^{-4}$ MeV) in the 14$^{\text{th}}$-19$^{\text{th}}$ rows in Table \ref{tab:sumrule-sig-ps}, eventually the sum rule is reproduced precisely with negligible errors
($\tilde V_{\rm sum} \lesssim 10^{-9}$ MeV). As depicted in Table \ref{tab:sumrule-sig-ps}, similar systematics in describing the sum rule, that the results are not shown here, are found with the relativistic formalism for the tensor force components in the Fock diagram of $\omega$-V, $\rho$-V, $\rho$-T and $\pi$-PV couplings. Besides, it is also found in the last six rows of Table \ref{tab:sumrule-sig-ps} that the relations (\ref{equ:sumrule2}) for pseudo-spin partners are fulfilled precisely if one neglects the upper components and takes the same radial wave functions for the lower components.

\begin{center}\setlength{\tabcolsep}{11pt}
\tabcaption{ \label{tab:sumrule-sig-ps}  Interaction matrix elements $V^T_{\tilde j_\pm \tilde j'}$ between the pseudo-spin partner states, namely the pseudo orbital $\tilde p$ and $\tilde d$ of $^{90}$Zr, for the tensor force components in the Fock diagram of the $\sigma$-S couplings. The 2$^{\text{nd}}$-7$^{\text{th}}$ rows shows the results (in units of $10^{-1}$ MeV) calculated with the radial wave functions determined by the self-consistent calculations of DDRHF with PKA1. For the results in the 8$^{\text{th}}$-13$^{\text{th}}$ rows the upper components in both the $\tilde j_\pm$ and the $\tilde j'_\pm$ orbits are omitted, and for ones in the 14$^{\text{th}}$-19$^{\text{th}}$ rows the $\tilde j_\pm$ orbits, as well as $\tilde j'_\pm$ orbits, share the same radial wave functions in addition to neglecting the upper components, and both are in units of $10^{-4}$ MeV.}
\footnotesize
\begin{tabular}{c|rrrr}\hline\hline
\multirow{2}{*}{$\tilde j_\pm$} & \multicolumn{4}{c}{$\tilde j'$} \\ \cline{2-5}
                     &$\nu1\tilde p_{3/2}$&$\nu 1\tilde p_{1/2}$&$\nu1\tilde d_{5/2}$&$\nu1\tilde d_{3/2}$  \\\hline
$\nu1\tilde{p}_{3/2}$&         0.929      &      $-$0.142       &           0.672    &        $-$0.400      \\
$\nu1\tilde{p}_{1/2}$&      $-$0.142      &      $-$0.460       &        $-$0.128    &        $-$0.396      \\
$\tilde V_{\rm sum}$ &         3.433      &      $-$1.487       &           2.432    &        $-$2.390      \\\hline
$\nu1\tilde{d}_{5/2}$&         0.672      &      $-$0.128       &           0.730    &        $-$0.272      \\
$\nu1\tilde{d}_{3/2}$&      $-$0.400      &      $-$0.396       &        $-$0.272    &           0.020      \\
$\tilde V_{\rm sum}$ &         2.432      &      $-$2.351       &           3.290    &        $-$1.555      \\\hline\hline
$\nu1\tilde{p}_{3/2}$&         0.931      &      $-$1.513       &           0.970    &        $-$1.161      \\
$\nu1\tilde{p}_{1/2}$&      $-$1.513      &         2.472       &        $-$1.578    &           1.902      \\
$\tilde V_{\rm sum}$ &         0.698      &      $-$1.106       &           0.722    &        $-$0.842      \\\hline
$\nu1\tilde{d}_{5/2}$&         0.970      &      $-$1.578       &           1.591    &        $-$1.872      \\
$\nu1\tilde{d}_{3/2}$&      $-$1.161      &         1.902       &        $-$1.872    &           2.229      \\
$\tilde V_{\rm sum}$ &         1.172      &      $-$1.863       &           2.057    &        $-$2.315      \\\hline\hline
$\nu1\tilde{p}_{3/2}$&         0.618      &      $-$1.236       &           0.634    &        $-$0.951      \\
$\nu1\tilde{p}_{1/2}$&      $-$1.236      &         2.472       &        $-$1.268    &           1.902      \\
$\tilde V_{\rm sum}$ &      $-$0.000      &         0.000       &        $-$0.000    &        $-$0.000      \\\hline
$\nu1\tilde{d}_{5/2}$&         0.634      &      $-$1.268       &           0.991    &        $-$1.486      \\
$\nu1\tilde{d}_{3/2}$&      $-$0.951      &         1.902       &        $-$1.486    &           2.229      \\
$\tilde V_{\rm sum}$ &         0.000      &      $-$0.000       &           0.000    &           0.000      \\\hline\hline
  \end{tabular}
\end{center}

In order to understand the systematics on describing the sum rule for pseudo-spin partners, Figure \ref{fig:Zr90wave} shows the radial wave functions $G(r)$ and $F(r)$ as functions of radial distance $r$. It can be seen that the radial wave functions $G(r)$ for the upper components are quite different for the pseudo-spin partners, from which the distinct violation of the sum rule can be well understood. On the contrary, as shown in the lower panels of Fig. \ref{fig:Zr90wave}, the lower components for the pseudo-spin partners are of similar radial dependence which account for the approximate PSS. It is also interesting to seen that the radial wave functions $F(r)$ for the pseudo-spin partners are quantitatively different at large radial distance. As a result, the sum rule for the pseudo-spin partners, if neglecting the upper components, is not fulfilled as good as the cases for the spin partners neglecting the lower components. Since the radial wave functions $F(r)$ are near one order of magnitude smaller than the upper ones $G(r)$, it is not surprising to see that the interaction matrix elements $V^T_{\tilde j\tilde j'}$ are reduced by 3 order of magnitude, if neglecting the upper components of nucleon spinors.

It should be mentioned that within the non-relativistic scheme \cite{Otsuka2005PRL95.232502} the sum rule (\ref{equ:sumrule}), as well as the relations (\ref{equ:sumrule2}), can be properly fulfilled, particularly if one adopts the same radial wave functions for the spin partners. While for the pseudo-spin partners, as depicted in the 2$^{\text{nd}}$-7$^{\text{th}}$ rows of Table \ref{tab:sumrule-sig-ps}, the relations (\ref{equ:sumrule}) and (\ref{equ:sumrule2}) shall not be satisfied any more due to the nodal difference. One may argue that with the identical wave functions these relations can be fulfilled. However, such arbitrary approximation already change the pseudo-spin doublet states themselves. In the relativistic scheme, the relations (\ref{equ:sumrule}) and (\ref{equ:sumrule2}) can be fulfilled properly for the spin partners if neglecting the lower components of nucleon spinors, and can be reproduced precisely if one further lets the spin partners share the same radial wave functions $G(r)$. Moreover, for the pseudo-spin partners, the sum rules can be also fulfilled to certain extent if one neglects the upper components, and further these relations can be satisfied precisely with identical radial wave functions $F(r)$ that corresponds to the exact PSS.

\begin{center}
\includegraphics[width=8cm]{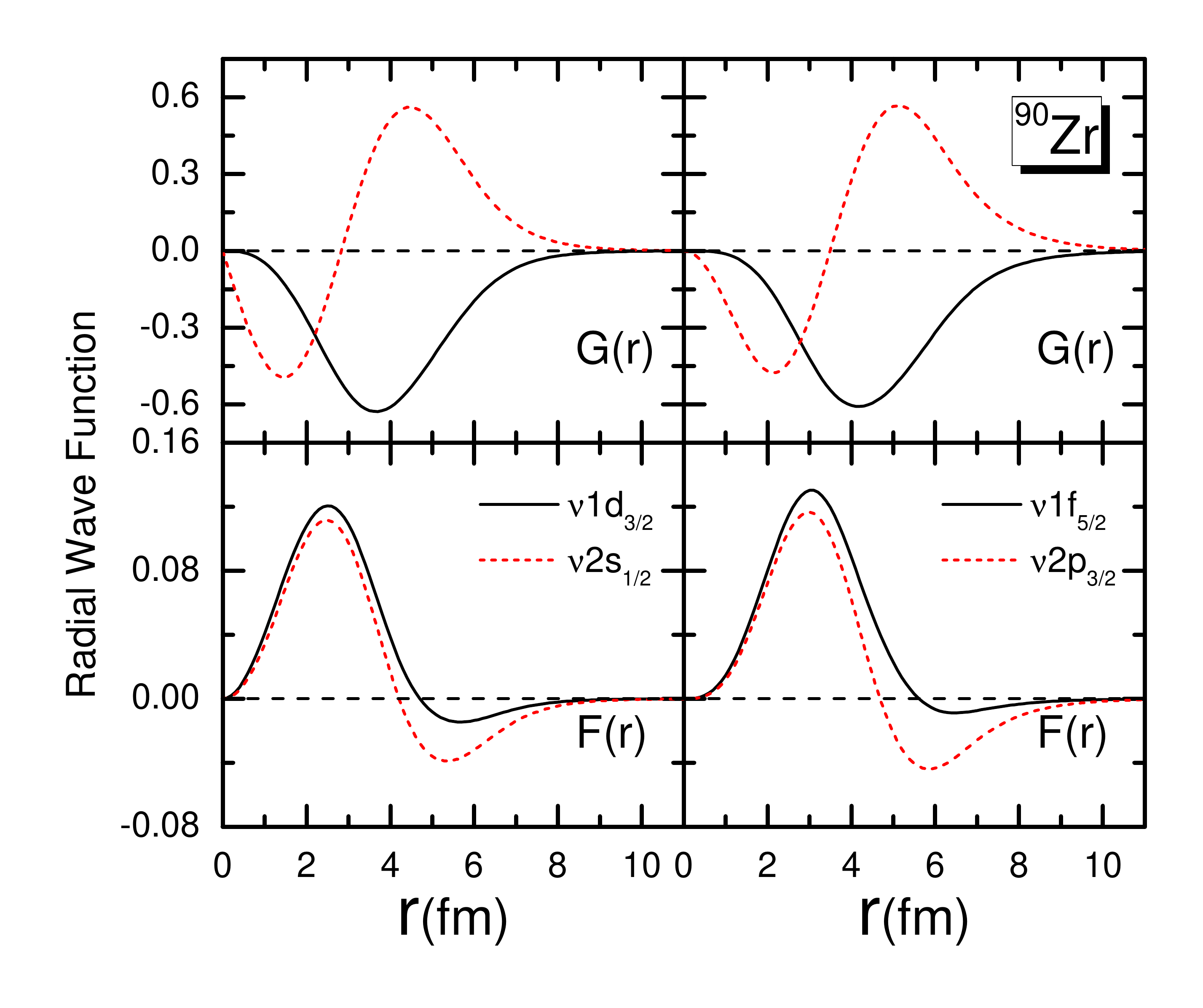}
\figcaption{\label{fig:Zr90wave}   (Color Online) Radial wave functions of the pseudo-spin partner states $\nu1\tilde{p}$ and $\nu1\tilde{d}$ in $^{90}$Zr, where $G(r)$ denotes the upper components and $F(r)$ corresponds to the lower components. The results are extracted from the calculations of DDRHF with PKA1.}
\end{center}

\subsection{Sum Rule for the Nodal States}

As discussed in previous subsections, the nodal differences on the lower components of spin-partners, as well as on the upper components for the pseudo-spin partners, induce distinct violation on the sum rules (\ref{equ:sumrule}) and (\ref{equ:sumrule2}). It is then worthwhile to check the validity of the sum rules on the nodal states with the relativistic formalism for the tensor force components in the Fock diagrams.

Taking the nodeless neutron orbit $\nu1h$, and the nodal ones $\nu2d$ and $\nu3p$ in $^{208}$Pb as the examples, Tables \ref{tab:sumrule-sig-nodes1}-\ref{tab:sumrule-sig-nodes3} shows the interaction matrix elements $V^T_{j_\pm j'}$ (in units of $10^{-2}$ MeV), calculated with the supplemented relativistic formalism for the tensor force components in the Fock diagram of $\sigma$-S coupling. As expected, the magnitude of $V^T_{j_\pm j'}$ tends to be smaller with the increasing of mass numbers, i.e., the units changing from $10^{-1}$ MeV to $10^{-2}$ MeV. Similar as the results in the 2$^{\text{nd}}$ to the 7$^{\text{th}}$ rows in Table \ref{tab:sumrule-sig}, Table \ref{tab:sumrule-sig-nodes1} shows the results calculated with the full nucleon spinors, and distinct deviations from the sum rule can be seen with the $V_{\rm sum}$ values comparable to $V^T_{j_\pm j'}$ themselves. Notice that the sum rules are scaled with the occupations $\hat j^2_\pm$, see Eqs. (\ref{equ:sumrule}) and (\ref{equ:sumrule2}), and as a result the violation of the sum rules is then enlarged with the increasing of angular momentum, particularly for the high-$j$ orbits $\nu1h$ and $\nu2f$. While for $V^T_{j_\pm j'}$ between the $\nu3p$ orbits, the deviations from the sum rule are also surprisingly large.

\begin{center}\setlength{\tabcolsep}{4pt}
\tabcaption{ \label{tab:sumrule-sig-nodes1}  Interaction matrix elements $V^T_{j_\pm j'}$ ($10^{-2}$ MeV) between the spin partner states, namely the nodeless neutron orbit $1h$, the nodal neutron orbits $2d$ and $3p$ of $^{208}$Pb, for the tensor force components in the Fock diagram of the $\sigma$-S couplings. The results are calculated with the radial wave functions determined by the self-consistent calculations of DDRHF with PKA1.}
\footnotesize
\begin{tabular}{c|rrrrrr}\hline\hline
\multirow{2}{*}{$j_\pm$} & \multicolumn{6}{c}{$j'$} \\ \cline{2-7}
              & $\nu1h_{11/2}$& $\nu1h_{9/2}$& $\nu2f_{7/2}$& $\nu2f_{5/2}$& $\nu3p_{3/2}$& $\nu3p_{1/2}$ \\ \hline
$\nu1h_{11/2}$&       3.048   &  $-$4.884    &      0.617   &    $-$1.603  &   $-$0.089   &     $-$0.713  \\
$\nu1h_{9/2}$ &    $-$4.884   &     3.673    &   $-$1.971   &       0.918  &   $-$0.778   &        0.021  \\
$V_{\rm{sum}}$&    $-$12.26   &  $-$21.88    &   $-$12.31   &    $-$10.05  &   $-$8.850   &     $-$8.354  \\ \hline
$\nu2f_{7/2}$ &       0.617   &  $-$1.971    &      1.114   &    $-$3.344  &   $-$0.231   &     $-$1.298  \\
$\nu2f_{5/2}$ &    $-$1.603   &     0.918    &   $-$3.344   &       2.186  &   $-$1.237   &        0.187  \\
$V_{\rm{sum}}$&    $-$4.680   &  $-$10.26    &   $-$11.15   &    $-$13.63  &   $-$9.274   &     $-$9.266  \\ \hline
$\nu3p_{3/2}$ &    $-$0.089   &  $-$0.778    &   $-$0.231   &    $-$1.237  &   $-$0.007   &     $-$4.389  \\
$\nu3p_{1/2}$ &    $-$0.713   &     0.021    &   $-$1.298   &       0.187  &   $-$4.389   &        3.600  \\
$V_{\rm{sum}}$&    $-$1.782   &  $-$3.072    &   $-$3.521   &    $-$4.577  &   $-$8.806   &     $-$10.35  \\
 \hline\hline
 \end{tabular}
\end{center}

Furthermore, if neglecting the lower components of nucleon spinors, i.e., the results in Table \ref{tab:sumrule-sig-nodes2}, the agreements with the sum rule are distinctly improved, despite the large scaling factors in high-$j$ orbits. Specifically, for $V^T_{j_\pm j'}$ between the $\nu3p$ orbits, the deviations from the sum rule tend to vanish. In fact it can also be found that for the repulsive-type $V^T_{j_\pm j'}$, namely $V^T_{j_+j_+'}$ and $V^T_{j_-j'_-}$,  the contributions from the lower component are also remarkable, which even change the sign of the interaction matrix elements (See Table \ref{tab:sumrule-sig-nodes1}). One may also notice that the $V_{\text{sum}}$ values for the high-$j$ orbits $\nu1h$ and $\nu2f$ are in general larger than the ones for $\nu3p$ orbits, see Table \ref{tab:sumrule-sig-nodes2}. It can be well understood from the radial wave functions $G(r)$ shown in Fig. \ref{fig:Pb208pfh}. It is shown that the difference on the wave functions $G(r)$ of the spin partners $\nu1h$ are remarkable due to the fact that the $\nu1h$ orbits cross over the major shell $N=82$, whereas the $\nu3p$ orbits have almost identical $G(r)$. If one lets the $j_+$ and $j_-$ orbits, as well as $j'_+$ and $j'_-$ orbits, share the same radial wave functions in addition to neglecting the lower components, as shown in Table \ref{tab:sumrule-sig-nodes3}, the sum rules can be reproduced precisely with negligible errors ($V_{\rm sum} \lesssim 10^{-7}$ MeV).

\begin{center}\setlength{\tabcolsep}{4pt}
\tabcaption{ \label{tab:sumrule-sig-nodes2}  Similar to Table \ref{tab:sumrule-sig-nodes1},  but the lower components of nucleon spinors in both the $j_\pm$ and the $j'_\pm$ orbits are omitted. }
\footnotesize
\begin{tabular}{c|rrrrrr}\hline\hline
\multirow{2}{*}{$j_\pm$} & \multicolumn{6}{c}{$j'$} \\ \cline{2-7}
              &$\nu1h_{11/2}$& $\nu1h_{9/2}$& $\nu2f_{7/2}$& $\nu2f_{5/2}$& $\nu3p_{3/2}$& $\nu3p_{1/2}$ \\ \hline
$\nu1h_{11/2}$&      3.820   &  $-$4.553    &      1.086   &    $-$1.384  &      0.260   &     $-$0.503  \\
$\nu1h_{9/2}$ &   $-$4.553   &     5.516    &   $-$1.453   &       1.807  &   $-$0.339   &        0.655  \\
$V_{\rm{sum}}$&      0.317   &     0.529    &   $-$1.504   &       1.463  &   $-$0.268   &        0.509  \\ \hline
$\nu2f_{7/2}$ &      1.086   &  $-$1.453    &      2.263   &    $-$2.880  &      0.462   &     $-$0.889  \\
$\nu2f_{5/2}$ &   $-$1.384   &     1.807    &   $-$2.880   &       3.703  &   $-$0.611   &        1.173  \\
$V_{\rm{sum}}$&      0.383   &  $-$0.786    &      0.828   &    $-$0.817  &      0.033   &     $-$0.072  \\ \hline
$\nu3p_{3/2}$ &      0.260   &  $-$0.339    &      0.462   &    $-$0.611  &      1.818   &     $-$3.533  \\
$\nu3p_{1/2}$ &   $-$0.503   &     0.655    &   $-$0.889   &       1.173  &   $-$3.533   &        6.872  \\
$V_{\rm{sum}}$&      0.033   &  $-$0.045    &      0.070   &    $-$0.095  &      0.204   &     $-$0.389  \\ \hline\hline
 \end{tabular}
\end{center}

\begin{center}\setlength{\tabcolsep}{4pt}
\tabcaption{ \label{tab:sumrule-sig-nodes3}  Similar to Table \ref{tab:sumrule-sig-nodes1},  but the $j_\pm$ orbits, as well as $j'_\pm$ orbits, share the same radial wave functions in addition to neglecting the lower components of nucleon spinors.}
\footnotesize
\begin{tabular}{c|rrrrrr}\hline\hline
\multirow{2}{*}{$j_\pm$} & \multicolumn{6}{c}{$j'$} \\ \cline{2-7}
              &$\nu1h_{11/2}$& $\nu1h_{9/2}$& $\nu2f_{7/2}$& $\nu2f_{5/2}$& $\nu3p_{3/2}$& $\nu3p_{1/2}$\\ \hline
$\nu1h_{11/2}$&      3.820   &  $-$4.584    &      1.086   &    $-$1.448  &      0.260   &     $-$0.520  \\
$\nu1h_{9/2}$ &   $-$4.584   &     5.501    &   $-$1.303   &       1.737  &   $-$0.312   &        0.624  \\
$V_{\rm{sum}}$&      0.000   &     0.000    &      0.000   &    $-$0.000  &   $-$0.000   &     $-$0.000  \\ \hline
$\nu2f_{7/2}$ &      1.086   &  $-$1.303    &      2.263   &    $-$3.018  &      0.462   &     $-$0.924  \\
$\nu2f_{5/2}$ &   $-$1.448   &     1.737    &   $-$3.018   &       4.024  &   $-$0.616   &        1.232  \\
$V_{\rm{sum}}$&      0.000   &     0.000    &      0.000   &       0.000  &   $-$0.000   &     $-$0.000  \\ \hline
$\nu3p_{3/2}$ &      0.260   &  $-$0.312    &      0.462   &    $-$0.616  &      1.818   &     $-$3.636  \\
$\nu3p_{1/2}$ &   $-$0.520   &     0.624    &   $-$0.924   &       1.232  &   $-$3.636   &        7.271  \\
$V_{\rm{sum}}$&   $-$0.000   &     0.000    &   $-$0.000   &    $-$0.000  &      0.000   &        0.000  \\ \hline\hline
\end{tabular}
\end{center}

Similar tests are also performed with the relativistic formalism for the tensor force components in the Fock diagrams of the $\omega$-V, $\rho$-V, $\pi$-PV and $\rho$-T couplings in DDRHF with PKA1, and the similar systematics are found for the nodal orbits on the tensor sum rule. Following the characteristics of tensor force, the interactions between the partners $\{j_\pm, j_+'\}$ are opposite to the ones between the partners $\{j_\pm,j'_-\}$. Notice that the states $j_\pm$ are often denoted with the quantum number $\kappa_\pm$, with $\kappa_\pm = \mp (j+1/2)$. In addition, with the relativistic representation of nucleon spinor, the signs of $\kappa$ are opposite for the upper and lower components. Thus, according to the spin-dependent feature of the tensor force, the signs of the $V^{T}_{G_jG_{j'}}$ and $V^{T}_{F_jF_{j'}}$ are different from the signs of the $V^{T}_{G_jF_{j'}}$ and $V^{T}_{F_jG_{j'}}$. So the dialog between the upper component of $j$ ($j'$) state and the lower one of $j'$ ($j$) state, namely $V^{T}_{G_jF_{j'}}$ ($V^{T}_{F_jG_{j'}}$), will cancel partly the dominant contributions of $V^{T}_{G_jG_{j'}}$ between the upper components of the states $j$ and $j'$ via the tensor force. As a result, distinct violation on the sum rules are then found in Tables \ref{tab:sumrule-sig}-\ref{tab:sumrule-sig-nodes3} due to the interference between the upper and lower components of nucleon spinors.

\begin{center}
\includegraphics[width=8cm]{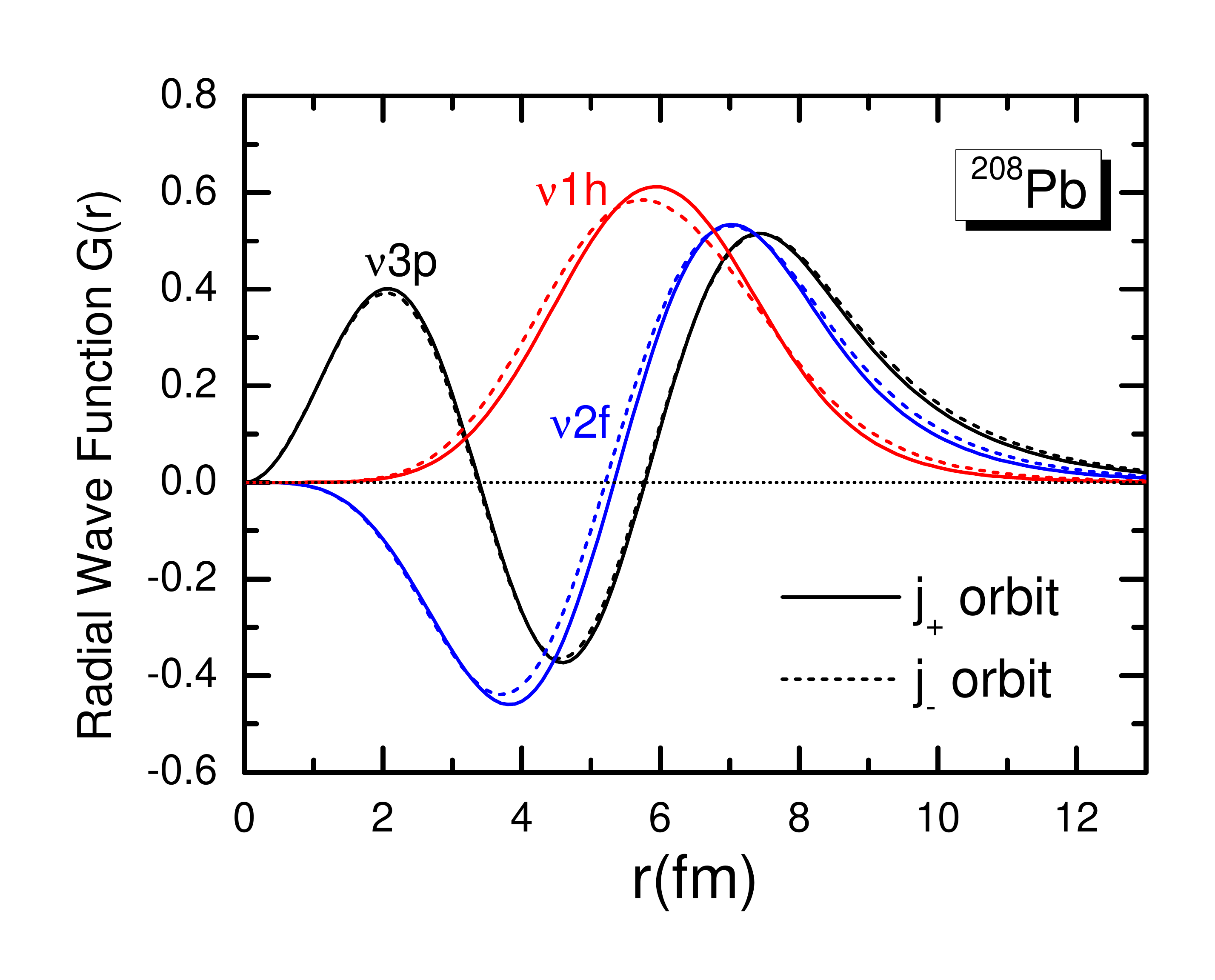}
\figcaption{\label{fig:Pb208pfh}   (Color Online) Radial wave functions $G(r)$ of the spin partner states $\nu1h$, $\nu2f$ and $\nu 3p$ in $^{208}$Pb. The results are extracted from the calculations of DDRHF with PKA1. }
\end{center}

\section{Summary}\label{sec:summary}

In conclusion, the nature of nuclear tensor force is illustrated in details within the relativistic Hartree-Fock approach, with a series of relativistic formalism of the tensor forces which are naturally introduced by the Fock diagrams of meson-nucleon couplings. Taking the original wave functions determined by the self-consistent DDRHF calculations, namely without dropping the lower component of Dirac spinors, the contributions to the spin-orbit splitting from the Fock diagrams are analyzed in selected realistic nuclei, and it is revealed that the spin-dependent feature described by the relativistic formalism for the tensor force is overestimated in the $\sigma$-S coupling channel. Drawing inspiration from different ranks of Lorentz tensor couplings used in various meson coupling channels, the relativistic formalism for the tensor force component in $\sigma$-S channel is then supplemented to its higher rank, which is proved to be impressive in fully interpreting the spin-dependent feature of the tensor force brought about by the Fock terms in realistic nuclei, without ignoring the lower components of nucleon spinors.

Taking the doubly magic nuclei $^{48}$Ca and $^{208}$Pb and the semi-magic one $^{90}$Zr as the candidates, the tensor sum rules are then tested for the spin and pseudo-spin partners with/without nodes, to further investigate the tensor force nature within relativistic model. Due to the opposite sign on $\kappa$ quantities of the upper and lower components, as well as the nodal difference, it is shown that the interference between two components of nucleon spinors brings distinct violations on the tensor sum rules in realistic nuclei. Even though, the spin dependence in the contributions to the spin-orbit splittings from the Fock diagrams can be almost fully taken into account by the supplemented relativistic formalism of the tensor force components. Moreover, if one neglects the lower/upper components of nucleon spinors for the spin/pseudo-spin partners, the sum rules can be fulfilled properly, and can be precisely reproduced if further taking the same radial wave functions for the spin/pseudo-spin partners, revealing the clear nature of the tensor force, illustrating the validity of the relativistic formalism of nuclear tensor forces as well.

\end{multicols}

\vspace{-1mm}
\centerline{\rule{80mm}{0.1pt}}
\vspace{2mm}


\begin{multicols}{2}

\end{multicols}

\clearpage

\end{document}